\documentclass[aps,prl,twocolumn,reprint,superscriptaddress]{revtex4-1}
\usepackage{graphicx}
\usepackage{amsmath}
\usepackage{amssymb}
\usepackage{colordvi}
\usepackage{mathrsfs}
\usepackage{bm}
\usepackage{verbatim}
\usepackage{dcolumn}
\usepackage{epsfig}
\usepackage{subfigure}
\usepackage[colorlinks,allcolors=blue]{hyperref}
\usepackage{ulem}
\usepackage{dsfont}
\usepackage{makecell}

\begin{document}
    \title{Chern Number Tunable Quantum Anomalous Hall Effect in Monolayer Transitional Metal Oxides via Manipulating Magnetization Orientation}
    \author{Zeyu Li}
	\affiliation{CAS Key Laboratory of Strongly-Coupled Quantum Matter Physics, and Department of Physics, University of Science and Technology of China, Hefei, Anhui 230026, China}
	\author{Yulei Han}
	\affiliation{Department of Physics, Fuzhou University, Fuzhou, Fujian 350108, China}
	\affiliation{CAS Key Laboratory of Strongly-Coupled Quantum Matter Physics, and Department of Physics, University of Science and Technology of China, Hefei, Anhui 230026, China}
	\author{Qian Niu}
	\affiliation{CAS Key Laboratory of Strongly-Coupled Quantum Matter Physics, and Department of Physics, University of Science and Technology of China, Hefei, Anhui 230026, China}
	\author{Zhenhua Qiao}
	\email[Correspondence author:~]{qiao@ustc.edu.cn}
	\affiliation{CAS Key Laboratory of Strongly-Coupled Quantum Matter Physics, and Department of Physics, University of Science and Technology of China, Hefei, Anhui 230026, China}
	\affiliation{ICQD, Hefei National Laboratory for Physical Sciences at Microscale, University of Science and Technology of China, Hefei, Anhui 230026, China}
	\date{\today}
	\begin{abstract}
      Although much effort has been made to explore quantum anomalous Hall effect (QAHE) in both theory and experiment, the QAHE systems with tunable Chern numbers are yet limited. Here, we theoretically propose that NiAsO$_3$ and PdSbO$_3$, monolayer transitional metal oxides, can realize QAHE with tunable Chern numbers via manipulating their magnetization orientations. When the magnetization lies in the \textit{x-y} plane and all mirror symmetries are broken, the low-Chern-number (i.e., $\mathcal{C}=\pm1$) phase emerges. When the magnetization exhibits non-zero \textit{z}-direction component, the system enters the high-Chern-number (i.e., $\mathcal{C}=\pm3$) phase, even in the presence of canted magnetization. The global band gap can approach the room-temperature energy scale in monolayer PdSbO$_3$ (23.4 meV), when the magnetization is aligned to \textit{z}-direction. By using Wannier-based tight-binding model, we establish the phase diagram of magnetization induced topological phase transition. Our work provides a high-temperature QAHE system with tunable Chern number for the practical electronic application.
    \end{abstract}

\maketitle

\textit{Introduction---.} As a typical representative of topological phases, the quantum anomalous Hall effect (QAHE) was initially theoretically predicted by Haldane in 1988~\cite{QHE,QAHE-Haldane}, being characterized by the quantized Hall conductance at zero magnetic field. Because of its dissipationless chiral edge states, the QAHE can be used to build next-generation low-power-consumption electronic devices, and provides a superior platform for investigating novel quantum phenomena such as topological superconductivity and topological magneto-electric effects~\cite{QAHEreview,Chiral-topo,Axion-th,Axion-hetro}. Therefore, the realization of QAHE becomes an extraordinarily important topic in the field of condensed matter physics. So far, numerous recipes have been proposed to realize the QAHE~\cite{Cr-Bi2Se3,Rashba-Graphene,Heavy-Atoms,Qiao-antiferromagnet,Chen-Fang,MnBi2Te4-1,MnBi2Te4-2,NiCl3,OsCl3,V2O3,Zheng-PtHgSe,moire1,moire2}. The first observation of QAHE is in Cr-doped (Bi, Sb)$_2$Te$_3$ thin films at a temperature of 30 mK~\cite{Exp-Cr-Bi2Se3}. Obviously, the low observation temperature is far away from the realistic application. Several works claimed that the magnetic dopants lead to magnetic inhomogeneity, which is detriment to the formation of QAHE and responsible for the low observation temperature~\cite{Imaging-doping1,Imaging-doping2,zheng-homo}. Therefore, the hope of raising the observation temperature becomes to find the intrinsic magnetic topological insulator. Fortunately, in a MnBi$_2$Te$_4$ thin film the QAHE was observed at the temperature of 1.4 K, which can further increase to 6.5 K when an external magnetic field is applied to drive the interlayer magnetic coupling from antiferromagnetic to ferromagnetic~\cite{Exp-MnBi2Te4}. This strongly indicates that the intrinsic magnetic insulator may breed high-temperature QAHE for practical application.

In addition to the temperature difficulty, tuning the Chern number to a higher one is another critical issue. High-Chern-number QAHE providing more dissipationless chiral edge channels can significantly improve the performance of quantum anomalous Hall devices. So far, several works have shown the theoretical possibility of QAHE with tunable Chern number in various magnetic topological insulator thin films, e.g., by increasing the thin film thickness or doping concentrations~\cite{jiang-high,wang-high,Doung-high,Zeng-high,xu-high}. The QAHE with tunable Chern number has been realized in thin MnBi$_2$Te$_4$ flakes or alternating magnetically doped topological insulator multilayer structures~\cite{high1,high2}. However, these Chern numbers are always fixed for any measured systems. Therefor, it is still challenging and interesting to explore material candidates that can realize QAHE with tunable Chern number for fixed system structures.

In this Letter, we systematically investigate the stability, magnetic, electronic and topological properties of monolayer transitional metal oxides NiAsO$_3$ and PdSbO$_3$ crystallized in $P$$\bar{3}$1$m$ space group by using first-principles calculations. These materials are predicted to be structurally stable by phonon spectra and molecular dynamics simulations. We show that they are \textit{x-y} easy-plane ferromagnetic half-metals with high Berezinskii-Kosterlitz-Thouless critical temperatures (i.e., 216.3 and 678.9~K for NiAsO$_3$ and PdSbO$_3$, respectively) and wide spin windows (1.26 and 0.97 eV). Six Dirac points appear in the first Brillouin zone due to $C_3$ and inversion symmetries. After considering spin-orbit coupling, the nontrivial band gaps open at the Dirac points to host QAHE. Remarkably, we find the Chern number $\mathcal{C}$ depends on the magnetization orientation, i.e., (i) when the magnetization lies in the \textit{x-y} plane and all mirror symmetries are broken, $\mathcal{C}=\pm1$; (ii) when the magnetization has a non-zero \textit{z}-component, high-Chern-number $\mathcal{C}=\pm3$ phase arises. We further find that the biaxial strain can efficiently tune the magnetocrystalline anisotropic energy (MAE), making it possible to change the magnetization orientation and realize Chern number tunable QAHE.

\begin{figure}
	\centering
	\includegraphics[width=0.45\textwidth]{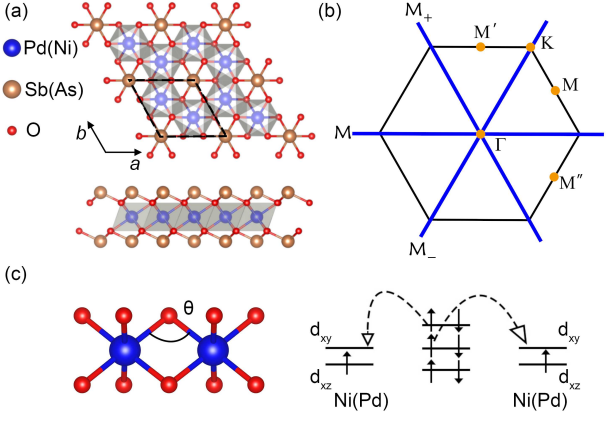}
	\caption{(a) Top and side views of monolayer NiAsO$_3$/PdSbO$_3$. (b) The first Brillouin zone with high symmetry points, the blue solid lines representing three vertical mirror planes. (c) The bond angle and the schematics of the Ni(Pd)-O-Ni(Pd) superexchange mechanism in the NiAsO$_3$ and PdSbO$_3$ via the (\textit{d$_{xz}$, d$_{yz}$})-\textit{p}-(\textit{d$_{xz}$, d$_{yz}$}) orbitals.}
	\label{fig1}
\end{figure}
\begin{table}
	\caption{Lattice constants, bond angle, binding energy, and total energy E for N\'eel, Stripe and Zigzag antiferromagnetic configurations per unit cell relative to the ferromagnetic ground state in meV.}
	\centering
	\renewcommand\arraystretch{2}
	\setlength\tabcolsep{1pt}
	\begin{tabular}{c|c|c|c|c|c|c|c|c}
		\hline\hline
		~ & $a$ (\r{A})  &  $c$ (\r{A})  &  $\theta$ ($^{\circ}$)  &  $E$$_b$ (eV)  &   N\'eel  &  stripe  &  zigzag \\\hline
		NiAsO$_3$     &     4.891     &     3.763     &     91.6     &     -1.062   &  62.7  &  436.4  &  173.2\\
		PdSbO$_3$     &     5.276     &     4.132     &     92.6     &     -0.999  &  93.3  &  213.4  &  65.6\\
		\hline\hline
	\end{tabular}
	\label{table1}
\end{table}

\textit{Structural Properties---.} The monolayer NiAsO$_3$ and PdSbO$_3$ share the same crystal structure with space group of $P$$\bar{3}$1$m$. As displayed in Fig.~\ref{fig1}(a), they include five atomic layers, with the transitional metal atom being surrounded by six oxygen atoms. The transitional metal atoms are in the oxidation state Ni$^{3+}$ and Pd$^{3+}$ with seven valence electrons, respectively. Their lattice constants are listed in Table~\ref{table1}. The structural stability is studied by using three different approaches. We first calculate the binding energy expressed as:
\begin{equation}
\begin{split}
E_{b} = E({\rm ABO_{3}}) - E({\rm A}) - E({\rm B}) - \frac{3}{2}E({\rm O_{2}}),\\
\end{split}
\end{equation}
where $E$(ABO$_3$), $E$(A), $E$(B) and $E$(O$_2$) are the total energies of monolayer NiAsO$_3$/PdSbO$_3$, Ni/Pd crystal, As/Sb crystal and oxygen molecule, respectively. The negative binding energy listed in Table~\ref{table1} indicates that its formation is an exothermic reaction. We then perform the phonon spectra calculation with 3$\times$3$\times$1 supercell, where the absence of imaginary phonon frequency strongly suggests the dynamical stability of monolayer NiAsO$_3$/PdSbO$_3$. Furthermore, we perform molecular dynamics simulation at 300 K, and the small fluctuations of total energy indicate their thermal stabilities at room temperature [see Fig. S1 in the Supplemental Material~\cite{Supplementary}]. In reality, the layered materials with the same structure at monolayer limit have been synthesized, e.g., SrRu$_2$O$_6$~\cite{SrRu2O6}, indicating the practical possibility of preparing monolayer NiAsO$_3$ and PdSbO$_3$.

\begin{figure}
  \centering
  \includegraphics[width=0.5\textwidth]{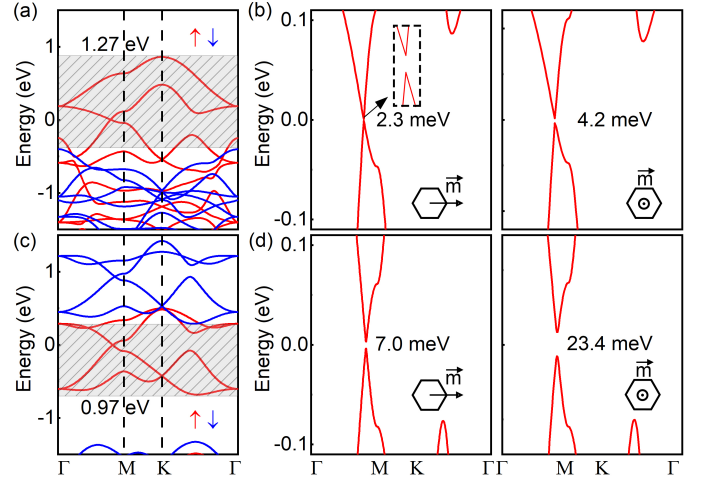}
  \caption{Spin-polarized band structure of monolayer NiAsO$_3$ (a) and PdSbO$_3$ (c). (b) Band structure of monolayer NiAsO$_3$ (b) and PdSbO$_3$ (d) with spin-orbit coupling when the magnetization along \textit{x}- and \textit{z}-direction, respectively.}
  \label{fig2}
\end{figure}
\begin{figure*}
  \centering
  \includegraphics[width=1.0\textwidth]{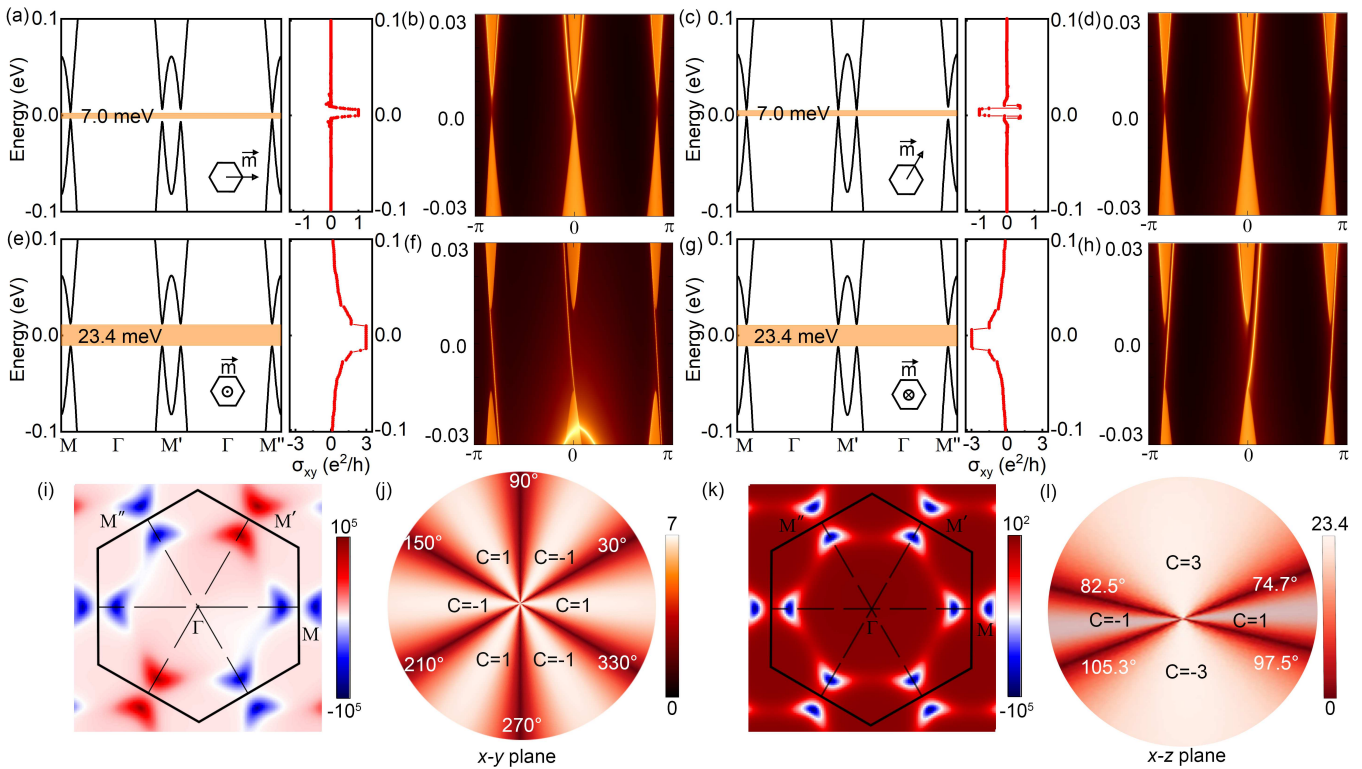}
  \caption{(a) and (c): Bulk band structure of monolayer PdSbO$_3$ along high symmetry line and corresponding anomalous Hall conductivity in the presence of spin-orbit coupling and magnetization lying in \textit{x-y} plane for different $\phi$= 0$^{\circ}$ (a) and $\phi$=60$^{\circ}$ (c), respectively. (b) and (d): Corresponding energy spectra of semi-infinite ribbon of monolayer PdSbO$_3$ for different $\phi$= 0$^{\circ}$ (b) and $\phi$=60$^{\circ}$ (d), respectively. Note that there is one gapless edge mode at k=0 in (b) and (d). (e)-(h) are similar to those in (a)-(d) but with magnetization along \textit{z}- and -\textit{z}-direction, respectively. Note that there are three gapless edge modes near k=0, $\pi$ and $-\pi$ in (f) and (h). (i) The distribution of the Berry curvatures for monolayer PdSbO$_3$ with in-plane magnetization along \textit{x}-direction and (k) out-of-plane magnetization along \textit{z}-direction, respectively. (i) Phase diagram of Chern number as a function of azimuthal angle $\phi$ for in-plane magnetization. (l) Phase diagram of Chern number as a function of polar angle $\theta$ for the out-of-plane magnetization.}
  \label{fig3}
\end{figure*}

\textit{Magnetic Properties---.} To reveal the magnetic ground state of monolayer NiAsO$_3$ and PdSbO$_3$, we consider four kinds of magnetic configurations: (i) ferromagnetic, (ii) N\'eel antiferromagnetic, (iii) stripe antiferromagnetic, and (iv) zigzag antiferromagnetic [see Fig. S2 ~\cite{Supplementary}]. The total energies listed in Table~\ref{table1} indicate that the ferromagnetic state is preferred in both NiAsO$_3$ and PdSbO$_3$. The nearest-neighbor transitional metal atoms are linked by oxygen atoms, implying that magnetic ground state is determined by the superexchange mechanism. The bond angle of Ni-O-Ni (Pd-O-Pd) is 91.6$^\circ$ (92.6$^\circ$) approaching 90$^\circ$, favoring the ferromagnetic interaction according to Goodenough-Kanamori rules~\cite{Goodenough,Kanamori}. The $d$ orbitals of transitional metal atoms surrounded by the distorted octahedra are split into three groups: \textit{d$_{z^2}$}, (\textit{d$_{xy}$ , d$_{x^2-y^2}$}), (\textit{d$_{xz}$ , d$_{yz}$})~\cite{RuI3,quart}. The projected band structures 	show that the six electrons first fill the \textit{d$_{z^2}$} and (\textit{d$_{xy}$ , d$_{x^2-y^2}$}) orbitals, then the remaining one occupies one of (\textit{d$_{xz}$ , d$_{yz}$}) orbitals with a net magnetic moment of 1 $\mu_B$. Because of the strong $pd\sigma$ hybridization between transitional metal atoms and oxygen atoms as displayed in Fig. S3 ~\cite{Supplementary}, the major exchange interaction is bridged via the orthogonal (\textit{p$_x$, p$_y$}) orbital of oxygen atom, resulting in the ferromagnetic coupling, as shown in Fig.~\ref{fig1}(c). The MAE illustrates that the magnetization lies in the \textit{x-y} plane, i.e., monolayer NiAsO$_3$ and PdSbO$_3$ belong to the category of $XY$ magnets [see Fig. S2 ~\cite{Supplementary}]. Although the Mermin-Wagner theorem prohibits the long-range magnetic order in two-dimensional isotropic systems, the finite-size effect can stabilize the magnetic order that has been recently verified in easy-plane magnet 1T-VSe$_2$ and CrCl$_3$ down to monolayer limit~\cite{finite,VSe2,CrCl3}. Our estimated critical Berezinskii-Kosterlitz-Thouless transition temperatures for monolayer NiAsO$_3$ and PdSbO$_3$ are 216.3 and 678.9 K, respectively~\cite{BKT1,BKT2}.

\textit{Band Structures and Topological Properties---.}
In Figs.~\ref{fig2}(a) and \ref{fig2}(c), one can observe that monolayer NiAsO$_3$ and PdSbO$_3$ are half-metals with wide spin windows (1.26 eV and 0.97 eV, respectively). The spin-down bands demonstrate a large energy gap around the Fermi level, whereas the spin-up bands exhibit metallic electronic structure, implying that they are promising candidates in spintronics such as spin valves and magnetic tunnel junctions~\cite{Half-Metallic Magnets,Spintronics}. Furthermore, along $\Gamma$-M high symmetry line, a spin-polarized Dirac point appears at the Fermi level. Due to the $C_3$ rotation and inversion symmetries, six spin-polarized Dirac points emerge in the first Brillouin zone.

The topological properties of monolayer NiAsO$_3$ and PdSbO$_3$ are investigated after turning on the spin-orbit coupling. Because their topological properties are similar we take monolayer PdSbO$_3$ as an example in the following part (see Fig. S6 for NiAsO$_3$ ~\cite{Supplementary}). Depending on the orientation of magnetization, various topological phases can be formed. As displayed in Fig.~\ref{fig1}(b), there are three mirror planes along $\Gamma$-K(K$^\prime$) high symmetry lines related to the $C_3$ rotation symmetry. When the magnetization is perpendicular to the mirror planes, the spin-polarized Dirac points are still preserved along the $\Gamma$-M line that parallel to the magnetization direction, leading to a zero Hall conductance [see Fig. S5 ~\cite{Supplementary}]. To obtain nonzero anomalous Hall conductivity with in-plane magnetization, it is crucial to break all mirror symmetries~\cite{inplane-QAHE1, inplane-QAHE2}. When the magnetization is along \textit{x}-direction, as shown in Figs.~\ref{fig3}(a)-~\ref{fig3}(b), a global gap about 7.0 meV opens in monolayer PdSbO$_3$ with a Chern number of $\mathcal{C}=1$. By tuning the orientation of in-plane magnetization, the system can be periodically driven into topological phases with alternative Chern numbers of $\mathcal{C}=\pm1$ in the interval of 60$^{\circ}$, e.g., $\mathcal{C}=-1$ phase appears when $\phi=60^\circ$ as displayed in Figs.~\ref{fig3}(c)-\ref{fig3}(d). Furthermore, the tiny MAE implies that an external magnetic field can be easily utilized to align magnetization deviating from the \textit{x-y} plane to \textit{z}-direction. As displayed in Figs.~\ref{fig3}(e)-\ref{fig3}(h), when the magnetization is aligned to $\pm$\textit{z}-direction,  the monolayer PdSbO$_3$ will enter a high-Chern-number ($\mathcal{C}=\pm3$) phase with a band gap increase to 23.4 meV.

We further construct a wannier-based tight-binding model by using (\textit{d$_{xz}$ , d$_{yz}$}) orbitals of Pd atom and \textit{p} orbitals of O atom to reveal the topological phase transition by tuning magnetization orientations. As shown in Fig. S5 ~\cite{Supplementary}, when the in-plane magnetization is perpendicular to the mirror plane, i.e., $\phi$ = 30$^{\circ}$ (210$^{\circ}$), 90$^{\circ}$ (270$^{\circ}$), 150$^{\circ}$ (330$^{\circ}$), the band gaps are closed with the crossing point located at the $\Gamma$-M (M$^{\prime}$, M$^{\prime\prime}$) high symmetry line, respectively. Therefore, the Chern number of the system alternatively changes between $\pm 1$ with a period of 60$^{\circ}$, as depicted in Fig.~\ref{fig3}(j). For the out-of-plane magnetization (e.g. \textit{x}-\textit{z} plane), starting from \textit{z}-direction (-\textit{z}-direction), the global band gap gradually decreases with the increase (decrease) of polar angle $\theta$. The gap will fully closed at $\theta$ = 74.7$^{\circ}$, 82.5$^{\circ}$, (97.5$^{\circ}$, 105.3$^{\circ}$), and then the gap reopens. The critical polar angle determines the phase boundaries of low- and high-Chern-number. Figure~\ref{fig3}(l) displays the phase diagram with different Chern numbers, where one can find that the high-Chern-number QAHE can even be realized with a canted magnetization.

The Berry curvatures can also help understand the Chern number variation. In Figs.~\ref{fig3}(i) and \ref{fig3}(k), one can clearly see the Berry curvature distribution for in-plane and out-of-plane magnetizations, respectively. For in-plane magnetization, two-thirds anticrossing zones contribute Berry curvatures with the same sign, but opposite in the remaining one-third zone, which is similar to that of LaCl~\cite{LaCl}. In contrast, for out-of-plane magnetization, the Berry curvatures give the same contribution at all anticrossing zones. Because each gapped Dirac point contributes $\pm\dfrac{1}{2}$ Chern number, the in-plane magnetization gives a small net value of low Chern number ($\mathcal{C}$=$\pm1$) but out-of-plane magnetization a high Chern number ($\mathcal{C}$=$\pm3$). 

\begin{figure}
  \centering
  \includegraphics[width=0.5\textwidth]{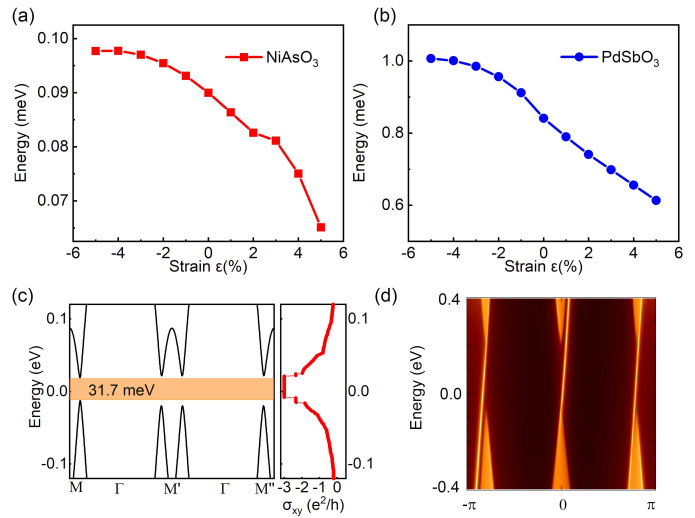}
  \caption{(a)-(b) Magnetocrystalline anisotropic energy of monolayer NiAsO$_3$ and PdSbO$_3$ as a function of applied biaxial strain. (c)-(d) Band structure along high symmetry lines, anomalous hall conductivity and energy spectra of semi-infinite ribbon of PdSbO$_3$-MoS$_2$ heterostructure with magnetization along \textit{z}-direction.}
  \label{fig4}
\end{figure}

\textit{Effect of Strain---.} Due to the flexibility of two-dimensional materials, applying strain can effectively tune various material properties~\cite{strain1,strain2,strain3,strain4,strain5}. We investigate the effect of strain on monolayer NiAsO$_3$ and PdSbO$_3$ by applying biaxial strain from -5$\%$ to 5$\%$. As displayed in Figs.~\ref{fig4}(a)-\ref{fig4}(b), the MAE decreases monotonically from compressive strain to tensile strain, implying that the tensile strain is beneficial to switch the QAHE between low- and high-Chern-number topological phases. It is known that the material growth on certain substrates naturally induces strain. For illustration, we construct a PdSbO$_3$-MoS$_2$ heterostructure (1 $\times$ 1 PdSbO$_3$ and $\sqrt3$ $\times$ $\sqrt3$ MoS$_2$), in which the PdSbO$_3$ is sandwiched by MoS$_2$ [see Fig. S8 ~\cite{Supplementary}]. After structural relaxation, the in-plane lattice constant of monolayer PdSbO$_3$ is enlarged by 3$\%$. The ferromagnetic state and the character of PdSbO$_3$ are still preserved in this heterostructure. For magnetization along \textit{z}-direction, the nontrivial band gap reaches 31.6 meV, exceeding the room-temperature energy scale [see Figs.~\ref{fig4}(c)-\ref{fig4}(d)].

\textit{Summary---.} We systematically investigate the electronic, and topological properties of the stable monolayer transitional metal oxides NiAsO$_3$ and PdSbO$_3$ based on the first-principles calculations. Their dynamical and thermal stabilities are confirmed by phonon calculations and molecular dynamics simulations. The magnetization orientation dependent QAHE with low- and high-Chern-number are proposed. The low-Chern-number QAHE is formed for the in-plane magnetization, while the high-Chern-number QAHE arises when the magnetization deviates from \textit{x-y} plane to \textit{z}-direction. Applying a tensile strain can effectively decrease the MAE in favor of the control for magnetization orientation by an external magnetic field. The tensile strain can be exerted by constructing a heterostructure accompanied by an increasing band gap in the phase with a high Chern number, such as in PdSbO$_3$-MoS$_2$ system. Our work provides an ideal platform to realize Chern number tunable QAHE for practical applications.

\textit{Acknowledgements---.} This work was financially supported by the National Natural Science Foundation of China (Grants No. 11974327 and No. 12004369), Fundamental Research Funds for the Central Universities (WK3510000010, WK2030020032), Anhui Initiative in Quantum Information Technologies (Grant No. AHY170000). We also thank the supercomputing service of AM-HPC and the Supercomputing Center of University of Scienceand Technology of China for providing the high performance computing resources.

\end{document}